\documentclass[11pt]{article}

\usepackage{amsfonts}
\usepackage{appendix}
\usepackage{verbatim}

\usepackage{amsmath}
\usepackage{amssymb}
\usepackage{latexsym}
\usepackage{graphicx}
\usepackage[english]{babel}
\usepackage[font={small}]{caption}
\usepackage{mathtools}
\usepackage{geometry}                
\usepackage{epstopdf}
\usepackage{tikz-cd}
\usetikzlibrary{cd}
\usepackage{amsfonts}
\usepackage{amssymb}
\usepackage{latexsym}
\usepackage{graphicx}
\usepackage[english]{babel}
\usepackage{tikz}
\usepackage{soul}
\allowdisplaybreaks

\usepackage{mciteplus}

\usepackage[colorlinks=true,      linkcolor=blue,      urlcolor=blue,      filecolor=blue,      citecolor=blue,
      pdfstartview=FitH,     pdfpagemode=UseNone,      bookmarksopen=true]{hyperref}  

\begin{document}
\input epsf

\def\p{\partial}
\def\h{{1\over 2}}
\def\be{\begin{equation}}
\def\bea{\begin{eqnarray}}
\def\ee{\end{equation}}
\def\eea{\end{eqnarray}}
\def\d{\partial}
\def\la{\lambda}
\def\eps{\epsilon}
\def\b{\bigskip}
\def\m{\medskip}

\renewcommand{\a}{\alpha}	
\renewcommand{\b}{\beta}
\newcommand{\g}{\gamma}		
\newcommand{\G}{\Gamma}
\renewcommand{\d}{\delta}
\newcommand{\D}{\Delta}
\renewcommand{\c}{\chi}			
\newcommand{\C}{\Chi}
\renewcommand{\P}{\Psi}
\newcommand{\s}{\sigma}		
\renewcommand{\S}{\Sigma}
\renewcommand{\t}{\tau}		
\newcommand{\e}{\epsilon}
\newcommand{\n}{\nu}
\renewcommand{\r}{\rho}
\renewcommand{\l}{\lambda}

\newcommand{\newsection}[1]{\section{#1} \setcounter{equation}{0}}

\def\q{\quad}

\def\h{{1\over 2}}
\def\t{\tilde}
\def\r{\rightarrow}
\def\nn{\nonumber\\}

\let\p=\partial

\newcommand\blfootnote[1]{%
  \begingroup
  \renewcommand\thefootnote{}\footnote{#1}%
  \addtocounter{footnote}{-1}%
  \endgroup
}


\begin{flushright}
\end{flushright}
\vspace{20mm}
\begin{center}
{\LARGE 
A freely falling graviton in the D1D5 CFT }
\\
\vspace{18mm}
\textbf{Bin} \textbf{Guo}{\footnote{guo.1281@buckeyemail.osu.edu}}~\textbf{and} ~ \textbf{Shaun}~  \textbf{Hampton}{\footnote{shaun.hampton@ipht.fr}}
\\
\vspace{10mm}
${}^1$Department of Physics,\\ The Ohio State University,\\ Columbus,
OH 43210, USA\\ \vspace{8mm}

${}^2$Institut de Physique Th\'eorique,\\
	Universit\'e Paris-Saclay,
	CNRS, CEA, \\ 	Orme des Merisiers,\\ Gif-sur-Yvette, 91191 CEDEX, France  \\

\vspace{13mm}

\vspace{8mm}
\end{center}

\vspace{4mm}

\thispagestyle{empty}
\begin{abstract}

We study a freely falling graviton propagating in AdS in the context of the D1D5 CFT, where we introduce an interaction by turning on a deformation operator.
We start with one left and right moving boson in the CFT. After applying two deformation operators, the initial bosons split into three left moving and three right moving bosons. 
We compute the amplitude for various energies and extrapolate the result to the large energy region.
At early times, the amplitude is linear in time. 
This corresponds to an infalling graviton becoming redshifted in AdS.
At late times, the amplitude is periodic, which agrees with the fact that a freely falling graviton will not be thermalized.

\vspace{3mm}

\end{abstract}
\newpage

\setcounter{page}{1}

\numberwithin{equation}{section} 

\tableofcontents

\newpage

\section{Introduction}

In the AdS/CFT correspondence \cite{maldacena,gkp, witten}, a free falling particle in the AdS can be described by the time evolution after a local quench in the strongly coupled CFT \cite{quench}. The time evolution, including the evolution of the entanglement entropy and the expectation value of energy momentum tensor, can be well understood by the quasi-particle picture \cite{Cardy_1}: a left and a right moving quasi-particle are emmited from the point of the local quench and propagate freely through the system. 

The time evolution can also be understood by extrapolating the result from the weakly coupled region of the CFT to the strongly coupled region. In the weakly coupled CFT, the left and right moving particles from the local quench split into a cascade of lower energy particles. Let's consider a simple toy model \cite{susskind_2}. We start with one left and one right moving particle localized at the place of the local quench. They have wavelength $1/E$, where $E$ is energy injected by the local quench. At step one, the left and the right moving particles split into two left and two right moving particles with wavelength $2/E$. At step two, the two left and two right movers will split to four left and four right movers with wavelength $4/E$. After step $m$, we will have $2^m$ left and $2^m$ right moving particles with wavelength $2^m/E$. 

If the splitting is caused by a local operator in the CFT, the time for each step can be obtained from dimensional analysis. If the first step takes time $1/E$, the second step should take time $2/E$ and the step $m$ should take time $2^m/E$. In this picture, the wavelength of the particles occupy a region centered at the location of the local quench. This region grows linearly with time as the wavelengths of particles grow linearly with time. 
In the strongly coupled region of the CFT and including all higher order corrections, this linear growth becomes the free propagation of quasi-particles.

As proposed in \cite{susskind_2}, the size of operator can be defined as the number of particles and thus the wavelength of each particle in the above process. The linear growth of wavelength can be identified as the operator growth behavior, which is studied in the SYK model \cite{syk_1,syk_2} and the higher dimensional toy model \cite{susskind_2}. Further work on operator growth where the gravitational dual is a black hole background is given in \cite{og bh}. 

In this paper, we study the splitting process in the context of the deformed D1D5 CFT \cite{dmw} which we describe in more detail in the next section. It has been conjectured that there is a point in the moduli space of this CFT where the theory is free. This is called the `orbifold' point \cite{orbifold2}. For more details and results of computations at the `orbifold' point see \cite{dmw,Burrington:2012yq, cm, Larsen:1999uk, Arutyunov:1997gt,sv,lm1,lm2,Dei:2019iym}.  In \cite{chaos}, by investigating the out-of-time-ordered correlator (OTOC), it was shown that the free theory did not exhibit any chaotic behavior, a phenomenon which is related to operator growth. However, this non-chaotic behavior is expected at the orbifold point, where the theory is non-interacting. In this paper we deform away from the orbifold point by turning on a marginal deformation of the theory. We will see that adding an interaction is critical to producing the splitting effect. For several works involving the deformation operator see \cite{Burrington:2014yia,Carson:2016uwf,Carson:2016cjj,peet,Avery:2010er,hm,dissertation}. 

Instead of using the OTOC as a probe of operator growth, we will study the splitting of high energy particles into a cascade of lower energy particles as mentioned previously. To be specific, we will study the simplest splitting where a left and a right moving high energy bosonic mode splits into three left and three right moving bosonic modes. 
This splitting process was studied in \cite{hm,dissertation}. It was found that the leading order splitting is caused by two deformation operators. 
The computation involves mapping to a covering space to resolve the twist operator in the deformation operator \cite{Carson:2016uwf}. Due to this complication, the splitting of high energy particles cannot be studied easily \cite{hm,dissertation} and the nature of the splitting mechanism is not clear. 

In this paper, we will study the splitting numerically and extrapolate the result to high energies. 
We will find that at early times, i.e.\,\,$t\lesssim \pi$, the two deformation operators bind together to form an effective local operator.
Because the effective operator is local, the splitting naturally satisfies the dimensional analysis mentioned previously. Thus we can obtain a linear growth of the operator size.
At late times $\pi\lesssim t$, the splitting amplitude has periodicity $2\pi$. The effect of the perturbation generates oscillations in the splitting amplitude, but does not lead to a secular term where the amplitude continues to grow \cite{hm}. This agrees with the expectation that a freely moving graviton in AdS will not split into lower energy gravitons or be excited into stringy states.

We outline the paper as follows. In Section \ref{sec 2} we review the D1D5 CFT and the deformation operator. In Section \ref{sec 3} we describe the splitting process which we plan to investigate. In Section \ref{amp comp} we outline the computation of the splitting amplitude. In Section \ref{sec numerical} 
we numerically compute this splitting process for various initial energies and extrapolate the result to large initial energies. In Section \ref{sec operator growth} we will interpret the early time behavior as caused by an effective local operator. We also discuss the relation of the splitting process to the behavior of a freely falling graviton in AdS.
Finally, in section \ref{sec discussion} we discuss our results and outlook.

\section{The D1D5 CFT}\label{sec 2}
Here we review the main details of the D1D5 CFT at the orbifold point as well as some properties of the deformation operator which we use to perturb away from the orbifold point.


Compactify type IIB string theory as
\be
M_{9,1}\rightarrow M_{4,1}\times S^1\times T^4
\label{compact}
\ee
On $S^1$ we wrap $N_1$ D1 branes and on $S^1\times
T^4$ we wrap $N_5$ D5 branes. 
The low energy limit of this brane bound state gives a CFT on
the circle $S^1$.

It has been conjectured that in the moduli space there is a point called the orbifold point where the CFT is free \cite{orbifold2}. At this point the CFT is described by
a 1+1 dimensional sigma model. In the Euclidienized theory the base space is a cylinder spanned by the complex coordinate
\be\label{define w}
w = \tau + i\s
\ee
where
\be
\tau, \sigma: ~~~0\le \sigma<2\pi, ~~~-\infty<\tau<\infty
\label{cylinder coordinates}
\ee
The sigma model has a target space which is the `symmetrized product' of
$N_1N_5$ copies of $T^4$,
\be
(T^4)^{N_1N_5}/S_{N_1N_5}
\ee
where each copy of $T^4$ gives 4 bosonic excitations $X^1, X^2, X^3,
X^4$ and 4 fermionic excitations $\psi^1,
\psi^2, \psi^3, \psi^4$ for the left movers. Similarly, for the fermions, the right movers are $\bar\psi^1,
\bar\psi^2,\bar\psi^3,\bar\psi^4$. The fermions have two types of boundary conditions on the $\sigma$ circle:
antiperiodic (NS sector) or periodic (R sector). 
This theory has a total central charge 
\be
c=6 N_1N_5\equiv 6N
\ee
In this $S^N$ orbifold theory, there are twist sectors where $k$ copies of the CFT are linked together to give a single copy of the CFT on a circle of length $2\pi k$. We call each such set of linked copies a `component string'.

\subsection{Symmetries of the CFT}

The D1D5 CFT has $\mathcal{N}=4$ supersymmetry for both left and right movers.
For each of the ${\cal N} = 4$ algebras,  the internal R symmetry group is
$SU(2)$, which gives
a global symmetry group $SU(2)_L\times SU(2)_R$.  The
quantum numbers of these two $SU(2)$ groups are denoted by
\be
SU(2)_L: ~(j, j_3);~~~~~~~SU(2)_R: ~ (\bar j, \bar j_3)
\ee
Geometrically, this corresponds to
rotational symmetry along the 4 spatial directions of $M_{4,1}$, which is $SO(4)_E\simeq SU(2)_L\times SU(2)_R$, where the subscript $E$ denotes `external'. 
We use $SO(4)_I$, where $I$ stands for
`internal', to denote the $SO(4)$ symmetry along the $T^4$.
The compactification of the
torus breaks this symmetry. However, this $SO(4)_I\simeq SU(2)_1\times SU(2)_2$ symmetry still gives a useful way to organize fields at the orbifold point.
The spinor indices $\alpha, \bar\alpha$ are used for $SU(2)_L$ and $SU(2)_R$
respectively. The spinor indices $A, \dot A$ are used for for $SU(2)_1$ and
$SU(2)_2$ respectively.

We group the 4 real fermions in the left moving sector into complex
fermions $\psi^{\alpha A}$. Similarly for the right moving fermions, we have $\bar{\psi}^{\bar\alpha  A}$. The 4 bosons $X^i$ form a vector in the
$T^4$. This vector can be decomposed into the $(\h, \h)$  representation of $SU(2)_1\times SU(2)_2$, which gives  scalars $X_{A\dot A}$. 
The mode expansions for 
the left moving bosonic and fermionic excitations on a component string labeled by $i$ and with winding $k_{i}$ are
\bea\label{modes}
\alpha^{(i)}_{A \dot A,m}&=&\frac{1}{2\pi }\int_{\sigma=0}^{2\pi k_{i}}\p_{w}X_{A \dot A}(w)e^{mw}dw~~~~~~m=\frac{q}{k_{i}}\nn
d^{\alpha A (i)}_{m}&=&\frac{1}{2\pi i }\int_{\sigma=0}^{2\pi k_{i}}\psi^{\alpha A}(w)e^{mw}dw
~~~~~~m=\frac{q}{k_{i}} ~~(\text{R}), ~~~m=\frac{q+\tfrac{1}{2}}{k_{i}} ~~(\text{NS})
\eea
where $q$ is an integer. The R and NS denote the R sector and NS sector respectively. We also have the corresponding bosonic mode $\bar \alpha_{A\dot A,m}$ and fermionic mode $\bar d^{\bar\alpha A}_{m}$ for the right movers. 

The $\mathcal{N}=4$ superconformal symmetry mentioned above is generated by the operators $L_{n}, G^\alpha_{\dot A,r},
J^a_n$ for the left movers and $\bar L_{n}, \bar G^{\bar\alpha}_{\dot A,r}, \bar
J^a_n$ for the right movers. This is the so-called small $\mathcal{N}=4$ symmetry. The full symmetry is actually larger: it is the contracted large $\mathcal{N}=4$ superconformal symmetry \cite{mms,Sevrin:1988ew}. It contains the following four bosonic modes and four fermionic modes as extra generators for the left movers
\be\label{global}
\sum_{i}\alpha^{(i)}_{A \dot A,m} ~~~~~~~~\sum_{i}d^{\alpha A (i)}_{m}
\ee
and similarly for the right movers. These kinds of modes with a sum over all copies are called `global' modes. More about these `global' modes and the large $\mathcal{N}=4$ superconformal symmetry in our convention can be found in the appendix of \cite{lifting,Guo:2020gxm}.

\subsection{Deformation of the CFT}

The orbifold CFT describes the `free' point in moduli space. To move towards the supergravity description, we add a marginal deformation operator $D$, which has conformal dimensions $(h, \bar h)=(1,1)$, to the action \cite{Burrington:2014yia,Carson:2016uwf,peet,Avery:2010er}
\be\label{defor S}
S\r S+\lambda \int d^2 z D(z, \bar z)
\ee
In this paper we choose $D$ to be a singlet under all the symmetries at the orbifold point
\be\label{D 1/4}
D=\frac{1}{4}\epsilon^{\dot A\dot B}\epsilon_{\alpha\beta}\epsilon_{\bar\alpha \bar\beta} G^{\alpha}_{\dot A, -\h} \bar G^{\bar \alpha}_{\dot B, -\h} \sigma^{\beta \bar\beta}
\ee
where $\sigma^{\beta\bar\beta}$ is a twist operator of rank $2$ in the orbifold theory. The twist operator can join two component strings of the CFT with winding numbers $k_1$ and $k_2$ into a component string with winding number $k_1+k_2$. It can also break a component string with winding number $k_1+k_2$ into two component strings with winding number $k_1$ and $k_2$.
The twist operator in (\ref{D 1/4}) is defined as
\bea
\sigma = \sum_{i< j}\sigma_{ij}
\eea
where $\sigma_{ij}$ twists the two copies labelled by $i$ and $j$.
The operators $ G$ and $\bar G$ are the left and right moving supercharge operators at the orbifold point. 

We will study the large $N$ limit, where $N=N_1 N_5$.
Following \cite{Gomis:2002qi,gn}, 
we define $g$ as
\footnote{By matching the string spectrum in the PP-wave limit \cite{gn}, the coupling $\lambda$ in (\ref{defor S}) can be identified with the six-dimensional string coupling $g_6=g_s\sqrt{Q_5/Q_1}$. The radius of $AdS_3$ and $S^3$ is $(R_{AdS}/l_s)^2=g_6 \sqrt{N}$. The first inequality in (\ref{gravity point}) arises from the requirement that the AdS radius be much larger than the string length ($R_{AdS}/l_s\gg 1$) while the second follows from the requirement that the string coupling itself be small ($g_s\ll 1$).}
\bea\label{t coupling}
g\equiv \lambda\,N^{1/2}  
\eea
where the coupling $g$ plays the role of the 't Hooft coupling. The coupling $g$ should not be confused with the string coupling $g_s$.
Assuming $N_1\sim N_5$, the perturbative CFT is described by $g\ll 1$, while the parameter region describing D1D5 supergravity solution is
\be\label{gravity point}
1\ll g \ll \sqrt{N}
\ee
In the following, we will do the computation in the perturbative CFT region where $g\ll 1$ and then extrapolate the results to the gravity region where $g$ is large.

\section{The Splitting Process}\label{sec 3}

Here we discuss the types of processes we will be looking at. We start with the NS vacuum of $N$ singly wound strings given by
\be\label{NS N}
|\Omega\rangle=|0\rangle^{(1)}|\bar 0\rangle^{(1)}|0\rangle^{(2)}|\bar 0\rangle^{(2)}\dots |0\rangle^{(N)}|\bar 0\rangle^{(N)}
\ee
where $|0\rangle^{(i)}$ and $|\bar 0\rangle^{(i)}$ label the NS vacuum of the left and right movers for the $i$th copy of the singly wound strings. The dual spacetime is $AdS_3\times S^3\times T^4$. Consider sending a graviton into AdS. In the CFT, this corresponds to a state with a left and a right moving bosonic mode 
\be\label{process a bar a}
\sum_{i}\alpha^{(i)}_{-m}\bar \alpha^{(i)}_{-m}|\Omega\rangle
\ee
where we omit the charge indices $A$ and $\dot A$ to write the state schematically. Here $(i)$ labels the copy that the modes act on. Notice that because a physical state must be symmetric among all copies we have summed over the copy label $i$ in state (\ref{process a bar a}). The actual initial state will be a localized wave packet, which is a superposition of the above states with different energies.

\begin{figure}
\centering
        \includegraphics[width=13.5cm]{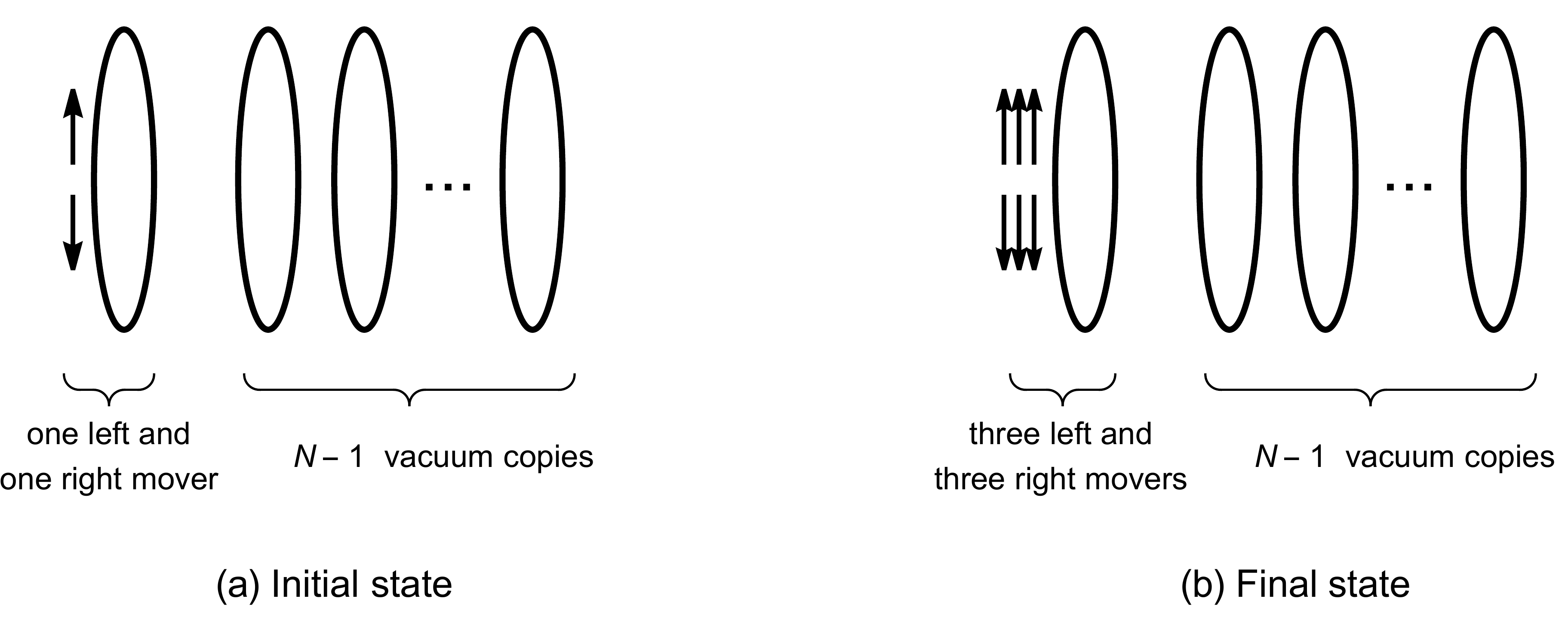}
\caption{The initial and final state of the $1\to3$ process. One left and one right mover split into three left and three right movers.}
\label{fig_splitting}
\end{figure}

At the orbifold point, in addition to a phase factor the initial state (\ref{process a bar a}) will not evolve with time since there is no interaction. Let us look at the evolution of the initial state when we deform the CFT away from the orbifold point. The initial state contains $N$ singly wound strings. The first deformation operator, which contains a twist operator, can twist any two singly wound strings into a doubly wound string. Then a second deformation operator can break the doubly wound string into two singly wound strings. After acting with two deformation operators, the one left and one right mover in the initial state can split into three left and three right movers. We call this a $1\to3$ process. This process is depicted in fig.\,\ref{fig_splitting}.

The left and right mover in each term of the initial state (\ref{process a bar a}) are on the same copy. Let us consider a more general process where the left and right movers can be on different copies
\be\label{process 1to3}
\alpha^{(i)}_{-m}\,\bar \alpha^{(j)}_{-m} |\Omega\rangle \rightarrow  (\alpha^{(i')}_{-m/3})^3\, (\bar \alpha^{(j')}_{-m/3})^3|\Omega\rangle
\ee
where $i,j,i',j'$ can only take two possible values from the range $1,\dots,N$ for any given amplitude. This is because the twist and untwist in this process only involves the same two copies.
In this paper, we consider the case that all three resulting left (right) movers are on the same copy $i'$ ($j'$).  
There can be more general cases where the resulting left (right) movers are on different copies but we will not consider them in this paper.
Throughout the process, the left and right movers split into more and more left and right movers with lower dimension due to the `collision' between left and right modes (\ref{process 1to3}).

The amplitude of the process is studied in \cite{hm}. The computation involves correlation functions with two twist operators. To compute this correlation function, we need to map it to the covering space and perform the Wick contractions. We outline this process in the next section. For the simplest case of $m=3$, the explicit result is written down in \cite{hm}. However, the complication of the computation grows  quickly with the dimension $m$. Due to this complication, no explicit result for general $m$ or even large $m$ is known. This makes the nature of the splitting (\ref{process 1to3}) unclear. In this paper, we will study the amplitude numerically upto $m=36$. We find that in the large $N$ limit, the initial and final bosonic modes turn to stay on the same component string. When the time scale is much larger than $1/m$, we find that the amplitude of the dominate process in the large $N$ limit is
\be
\tilde{A}^{ii\rightarrow ii}_m(t)\approx g^2 \frac{c}{m^2} [t]_{\text{saw-like}, 2\pi}~~~~~~~~~~m\rightarrow \infty
\ee
with $c\approx \frac{\pi}{18}$ and $i=1,\dots,N$. The $t$ is the time interval between the initial and final states and $g$ is the 't Hooft coupling (\ref{t coupling}). The `saw-like, $2\pi$' means that we start with the function in the square bracket in the region $(0,\pi)$ and then reflect it across $\pi$ to obtain the function in the region $(\pi,2\pi)$. Then we make a saw-like function with periodicity $2\pi$. For an exmaple, see fig.\,\ref{fig_amp}.

One may wonder if the following simpler process exists. Instead of having a `collision' between a left and right mover, we may consider the `splitting' of a left (right) mover only. We consider the initial state
\be\label{process a}
\sum_{i}\alpha^{(i)}_{-m}|\Omega\rangle
\ee
where we have summed the copy label $i$ to make the state symmetric among all copies. The bosonic mode in state (\ref{process a}) is the `global' mode mentioned in (\ref{global}). It is a generator of the symmetry, which is kept by the deformation operator. The deformation cannot split the bosonic mode in state (\ref{process a}). Thus the process (\ref{process 1to3}) is the simplest splitting process.

\section{Computing the Amplitude}\label{amp comp}

The method of computing the amplitude for splitting (\ref{process 1to3}) is studied in detail in \cite{dissertation}.
In this section, we will compute them numerically and find an empirical formula for large dimension $m$.
Without loss of generality, let us call the two copies involved in the process (\ref{process 1to3}) copy 1 and copy 2.
The NS vacuum in both the left and right moving sector for these two copies is
\be
|\Omega_{1,2}\rangle=|0\rangle^{(1)}|\bar 0\rangle^{(1)}|0\rangle^{(2)}|\bar 0\rangle^{(2)}
\ee
The effect of other copies in (\ref{NS N}) and the large $N$ limit will be studied in Section \ref{sec N}. Consider the following splitting process coming from the effect of two deformation operators
\bea\label{process}
|\Phi_0;i,j\rangle&=&\alpha^{(i)}_{--,-m}\bar \alpha^{(j)}_{++,-m} |\Omega_{1,2}\rangle\nn
\rightarrow |\Phi_f;i',j'\rangle&=&\alpha^{(i')}_{--,-p} \alpha^{(i')}_{++,-q} \alpha^{(i')}_{--,-r} 
~ \bar \alpha^{(j')}_{++,-p}  \bar \alpha^{(j')}_{--,-q}  \bar \alpha^{(j')}_{++,-r} |\Omega_{1,2}\rangle
\eea
where the copy labels $i,j,i',j'$ can be $1,2$. Here we consider a particular case of charge indices in the process (\ref{process 1to3}).
Next, we write down the amplitude that we'll need to compute.

\subsection{The Amplitude}

We put the initial state $|\Phi_0;i,j\rangle$ at time $0$ and the final state $|\Phi_f;i',j'\rangle$ at time $\tau$. In the time interval between them, there are two insertions of deformation operators. 
The explicit amplitude $1\to 3$ that we'd like to compute is given by 
\bea
A_n^{ij\rightarrow i'j'}(\tau)&\equiv&{1\over2}\lambda^2\int d^2w_2d^2w_1{1\over mpqr(1+\d_{p,r})} \langle \Phi_f;i',j'| D(w_2,\bar w_2) D(w_1,\bar w_1) |\Phi_0;i,j\rangle
\cr
\cr
&\equiv&{1\over2}\lambda^2\int d^2w_2d^2w_1\mathcal{A}^{ij\rightarrow i'j'}(w_1,w_2,\bar w_1, \bar w_2)
\label{amplitude}
\eea
where the $w_i$ is defined by
\be
w_i= \tau_i + i \sigma_i
\ee
and the region of integral is
\be
0\le \sigma_{i}<2\pi, ~~~0<\tau_{i}<\tau~~~~i=1,2
\ee
The amplitude before the integral is
\bea
&&\!\!\!\!\!\!\!\!\mathcal{A}^{ij\rightarrow i'j'}(w_1,w_2,\bar w_1, \bar w_2)\cr
&&\equiv \e^{\dot{C}\dot{D}}\e^{\dot{A}\dot{B}}{1\over mpqr(1+\d_{p,r})}\cr
&& {}^{(1)}\langle 0|{}^{(2)}\langle 0|\a^{(i')}_{++,p}\a^{(i')}_{--,q}\a^{(i')}_{++,r}G^+_{\dot{C},-{1\over2}}\s^-(w_2)G^-_{\dot{A},-{1\over2}}\s^+(w_1)\a^{(i)}_{--,-m}|0\rangle^{(1)}|0\rangle^{(2)}\cr
&&{}^{(1)}\langle \bar{0}|{}^{(2)}\langle \bar{0}|\bar{\a}^{(j')}_{--,p}\bar{\a}^{(j')}_{++,q}\bar{\a}^{(j')}_{--,r}\bar{G}^+_{\dot{D},-{1\over2}}\bar{\s}^-(\bar{w}_2)\bar{G}^-_{\dot{B},-{1\over2}}\bar{\s}^+(\bar{w}_1)\bar{\a}^{(j)}_{++,-m}|\bar{0}\rangle^{(1)}|\bar{0}\rangle^{(2)}
\eea
Here we have taken the conjugate of the final state $|\Phi_f;i',j'\rangle$.
We see a factorization between left and right movers. The front factor ${1\over mpqr(1+\d_{p,r})}$ comes from the normalization of initial and final states. We will not present the details of the computation of $\mathcal{A}$ here as they are lengthy but straightforward. They are provided in full detail in \cite{hm}. The outline of the steps of the computation are as follows
\begin{enumerate}
\item The amplitudes above are defined on a base space which has the geometry of a cylinder. The cylinder coordinate is given by $w$ which is expressed in (\ref{define w}), with a range given in (\ref{cylinder coordinates}).
The operator insertions on the cylinder contain twist operators which make the fields multivalued. To remove this ambiguity we map the cylinder amplitude to a covering space labeled by the coordinate $t$. The fields in the covering space are single valued. To do this we first map the cylinder to the complex $z$ plane through
\bea
z=e^w
\eea
and then to the covering space through
\bea
z = {(t+a)(t+b)\over t}
\eea
For details about the map see \cite{Carson:2016uwf}.

\item When mapping to the $t$-plane the amplitudes take the schematic form
\bea
A\to CA_t
\eea
The constant $C$ contains Jacobian factors coming from coordinate transformations from they $w$-cylinder, to the $t$-plane. The details of these transformations are given in \cite{dissertation,lifting}. In the $t$-plane the bosonic and fermionic fields are now single valued. There are also other various operator insertions which are present. The $t$-plane amplitude, $A_t$, arises from Wick contractions between pairs of bosonic and fermionic operators. The methods used to compute these contractions are given in  \cite{Carson:2016uwf}. They are combined to compute $A_t$ which is recorded in detail in \cite{dissertation}.
\item This computation yields both holomorphic and antiholomorphic parts of the amplitude which are finally multiplied together and integrated over the positions of the deformation insertions. This provides the full answer which we present numerically.
\end{enumerate}
After carrying out the above steps we find that $\mathcal{A}$ has the form
\bea
\mathcal{A}^{ij\rightarrow i'j'}(w_1,w_2,\bar w_1, \bar w_2) &=& \sum_{k=-(2m-1)\in\mathbb{Z}_{\text{odd}}}^{2m-1}~~\sum_{\bar k=-(2m-1)\in\mathbb{Z}_{\text{odd}}}^{2m-1}B^{ij\to i'j'}_{k,\bar k}(m;p,q,r) e^{{k \Delta w\over2}+{\bar k \Delta \bar w\over2}}\nn
\eea
where 
\bea
\Delta w = w_2 - w_1~~~~~\Delta \bar w = \bar w_2 - \bar w_1 
\eea
Here  $-k/2$ and  $-\bar k/2$ give the left and right dimensions (relative to the initial dimension) of the intermediate states in the region $\tau_1< \tau < \tau_2$ between the two deformation operators. Note that
$k$ and $\bar k$ have an upper bound and a lower bound. Thus only finite number of intermediate states contribute. Further, $k$ and $\bar k$ are odd integers.  

\subsection{Integrating the Amplitude}

In this section we integrate over the twist insertions to obtain the full amplitude. We begin with the expression
\bea\label{A integral}
A_m^{ij\rightarrow i'j'}(t)&=&{1\over2}\lambda^2\int d^2w_2\, d^2w_1\, \mathcal{A}^{ij\rightarrow i'j'}(w_1,w_2,\bar w_1,\bar w_2)\cr
&=&{1\over2}\lambda^2\int d^2w_2\,d^2w_1\,\sum_{k=-(2m-1)\in\mathbb{Z}_{\text{odd}}}^{2m-1}~~\sum_{\bar k=-(2m-1)\in\mathbb{Z}_{\text{odd}}}^{2m-1}B^{ij\to i'j'}_{k,\bar k}(m;p,q,r) e^{{k \Delta w\over2}+{\bar k \Delta \bar w\over2}}\nn
&\equiv&\lambda^2\sum_{k=-(2m-1)\in\mathbb{Z}_{\text{odd}}}^{2m-1}~~\sum_{\bar k=-(2m-1)\in\mathbb{Z}_{\text{odd}}}^{2m-1}B^{ij\to i'j'}_{k,\bar k}(m;p,q,r) I_{k,\bar k}
\eea
where
\bea\label{I}
I_{k,\bar k}&=& {1\over2}\int d^2w_2\,d^2w_1\, e^{{k \Delta w\over2}+{\bar k \Delta \bar w\over2}}
\eea
To obtain a real space amplitude we must wick rotate back to Lorentzian signature by taking $\tau \to it$. This gives
\bea
\Delta w &=& w_2 - w_1=i(t_2 - t_1 + \sigma_2 - \sigma_1)\cr
\Delta \bar w &=& \bar w_2 - \bar w_1 =i(t_2 - t_1 - (\sigma_2 - \sigma_1))
\eea
The region of integration is
\be
0\le \sigma_{i}<2\pi, ~~~0<t_{i}<t~~~~i=1,2
\ee
Because $k$ and $\bar k$ are odd numbers, $(k-\bar k)/2$ is an integer and $k,\bar k\neq 0$.
Due to the integration of $\sigma_i$, only terms with $k=\bar k$ contribute. 
Thus in (\ref{A integral}) only terms with
\be
k=\bar k \neq 0
\ee
contribute. In this case, we have
\bea
I_{k=\bar k \neq 0}(t)
&=&\frac{1}{2}\int d^2w_2\,d^2w_1\, e^{{k \Delta w\over2}+{\bar k \Delta \bar w\over2}}\cr
&=&\int_{-{t\over2}}^{{t\over2}}dt_2\int_{-{t\over2}}^{t_2}dt_1\int_{\s=0}^{2\pi}d\s_2\int_{\s=0}^{2\pi}d\s_1\, e^{i k (t_2-t_1)}\cr
&=& {4i \pi^2 \over k ^2} \bigg(kt  -  2e^{i {kt\over2}}\sin(  {kt\over 2})  \bigg)
\label{integral three}
\eea
Thus the integrated amplitude is
\bea\label{A final}
A_m^{ij\rightarrow i'j'}(t)=\lambda^2 \sum_{k=-(2m-1)\in\mathbb{Z}_{\text{odd}}}^{2m-1}~~~B^{ij\to i'j'}_{k, k}(m;p,q,r) {4i \pi^2 \over k ^2} \bigg(kt  -  2e^{i {kt\over2}}\sin(  {kt\over 2})  \bigg)
\eea
In \cite{hm}, it was found that 
\be\label{B sym}
B^{ij\to i'j'}_{k, k}(m;p,q,r) = B^{ij\to i'j'}_{-k, -k}(m;p,q,r)
\ee
Thus the integrated amplitude can be simplified to
\be\label{amp B}
A_m^{ij\rightarrow i'j'}(t)=\lambda^2 \sum_{k=1\in\mathbb{Z}_{\text{odd}}}^{2m-1}~~~B^{ij\to i'j'}_{k, k}(m;p,q,r) {16 \pi^2 \over k ^2} 
    \sin^2(  {kt\over 2}) 
\ee
We find that the terms linear in $t$ in (\ref{A final}) are cancelled. The integrated amplitude is periodic with periodicity $2\pi$.
Notice that to get this periodic behavior the following two properties are essential. Firstly, there is no intermediate states with exactly the same left and right dimensions as the initial state. Otherwise, from eq.\,(\ref{I}) there is a term 
\bea\label{t^2}
I_{k= \bar k = 0}(t)
=\frac{1}{2}\int d^2w_2\,d^2w_1 = 2\pi^2 t^2
\eea
This term is not periodic and grows with time.
Secondly, the intermediate states with energies increasing and decreasing by the same amount compared to the initial energy should have the same amplitude as shown in (\ref{B sym}).
In appendix \ref{B values}, we list the results of $B^{11\to 11}_{k, k}(m;m/3,m/3,m/3)$ for $m=18, 24, 30, 36$.

\section{Numerical Results}\label{sec numerical}

\subsection{The amplitude}\label{sec amp}

In section \ref{amp comp}, we review the computation of the amplitude $A_m^{ij\rightarrow i'j'}(t)$ (\ref{A integral}). There is no closed form for general $m$. In this section, we will study the large $m$ behavior. Let us start with $A_m^{11\rightarrow 11}(t)$, where the initial and final bosonic modes are placed in the copy 1 in the process (\ref{process}). 

\begin{figure}
\centering
        \includegraphics[width=7.6cm]{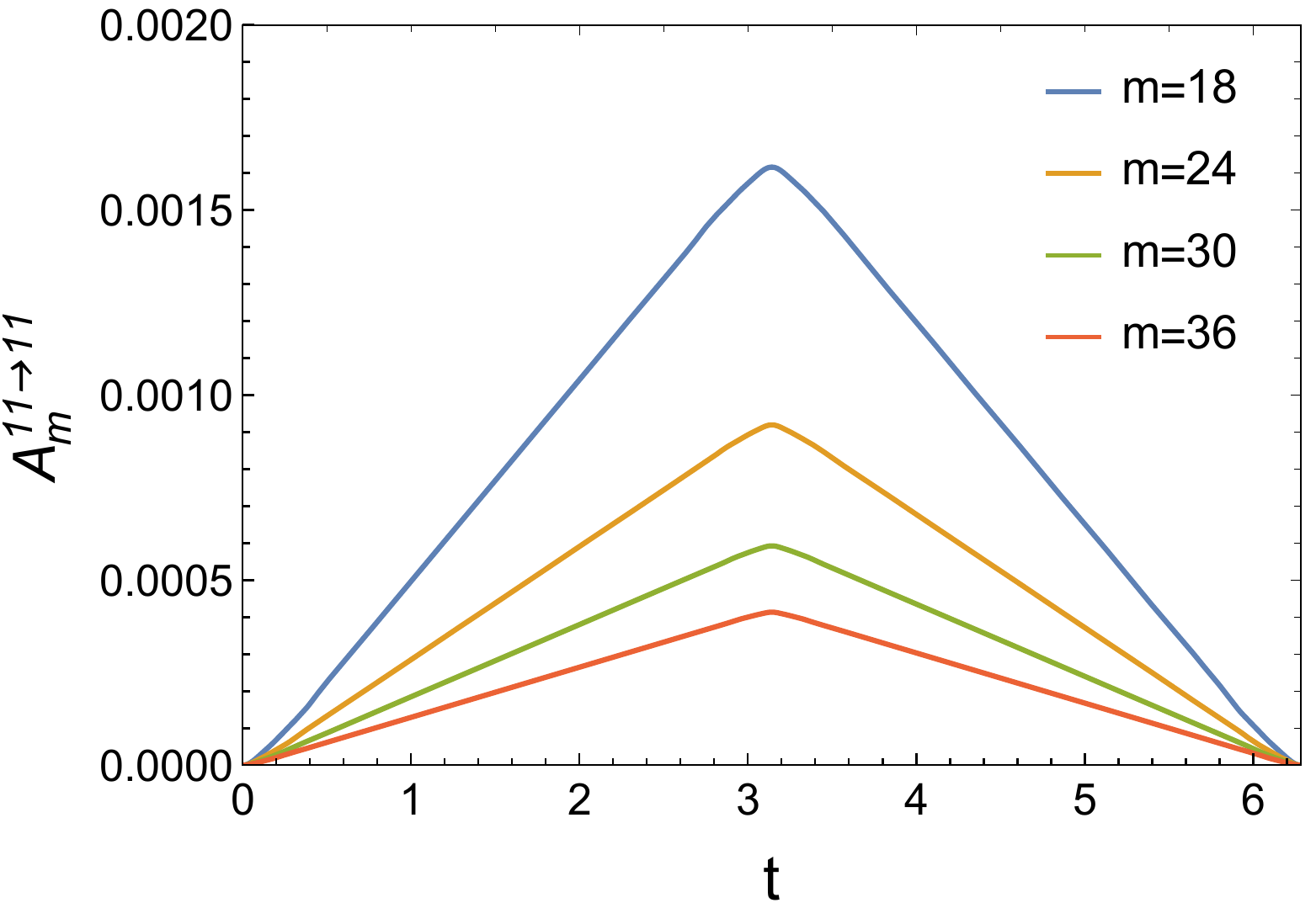}\quad
       \includegraphics[width=7.4cm]{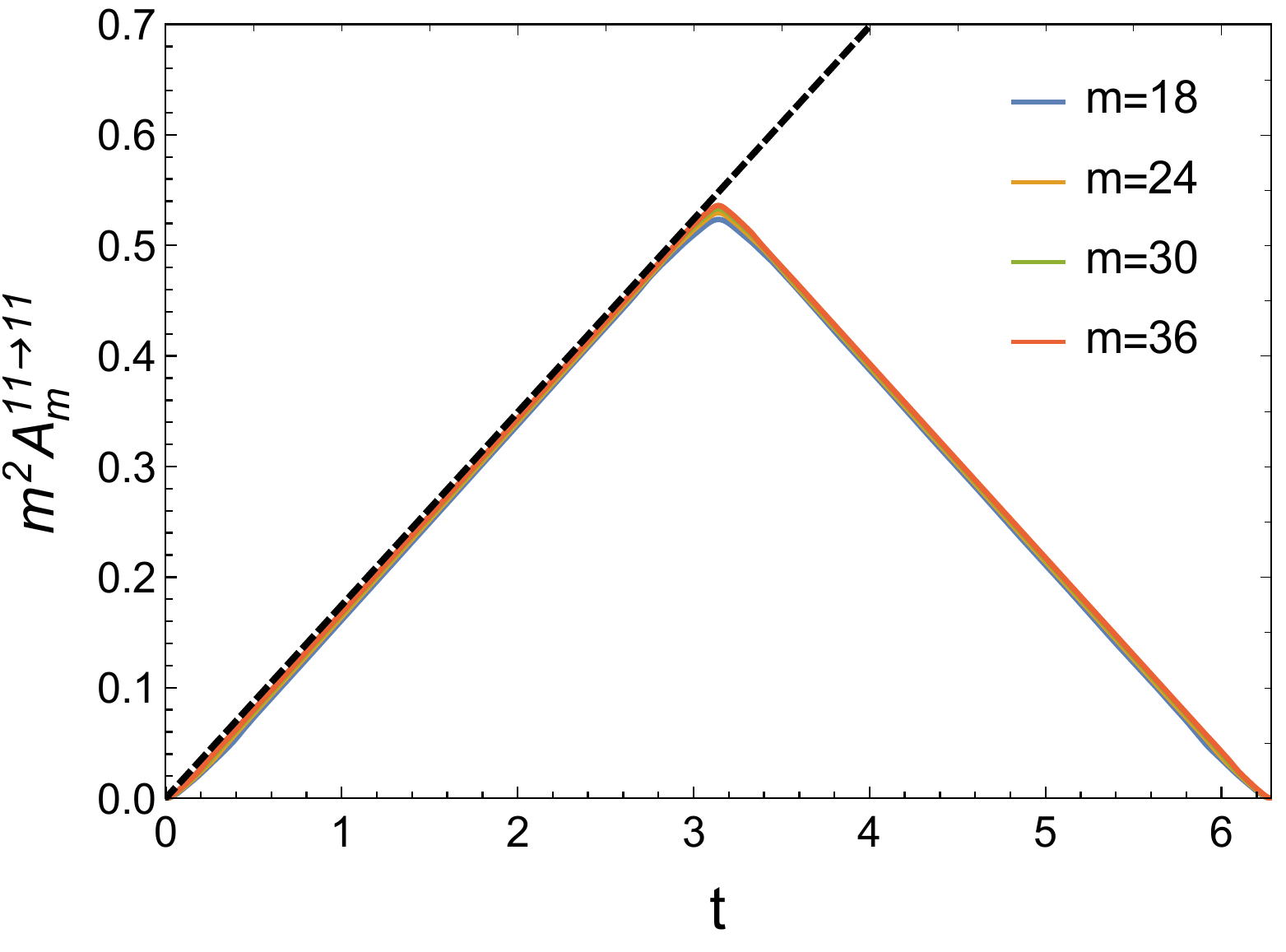}
\caption{Left: $A^{11\rightarrow 11}_{m}(t)$ for different $m$. Right: $m^2 A^{11\rightarrow 11}_{m}(t)$ for different $m$ and $\frac{\pi}{18}t$ (black dashed). In the plot, we set $\lambda^2=1$.}
\label{fig_amp}
\end{figure}

Using the numerical results in Appendix \ref{B values}, the amplitude $A_m^{11\rightarrow 11}(t)$ and the rescaled amplitude $m^2 A_m^{11\rightarrow 11}(t)$ as a function of $t$ for different $m$ are shown in fig.\,\ref{fig_amp}. From the left panel of fig.\,\ref{fig_amp}, we find that the amplitude is an approximate linear function of $t$ in the region $0\lesssim t \lesssim \pi$ for large enough $m$. The slopes are different for different $m$.    
To find the $m$ dependence, in the right panel of fig.\,\ref{fig_amp} we plot the rescaled amplitude $m^2 A_m^{11\rightarrow 11}(t)$. The slopes are approximately the same for different $m$. Thus we conclude for large enough $m$
\be\label{amp linear}
A_m^{11\rightarrow 11}(t) \approx \lambda^2 \frac{c}{m^2} t
\ee
where $c$ is a constant. In the right panel of fig.\,\ref{fig_amp}, we also plot the linear function $\frac{\pi}{18}t$ with the rescaled amplitude, which indicates
\be\label{slop const}
c\approx \frac{\pi}{18}
\ee
For the small $t$ region, by taking $\sin(kt/2)\approx kt/2$ in (\ref{amp B})
we have
\be\label{amp t2}
t\rightarrow 0~:~~~~~~~A_m^{11\rightarrow 11}(t)=\lambda^2 \sum_{k=1\in\mathbb{Z}_{\text{odd}}}^{2m-1}~B^{11\to 11}_{k, k}(m;p,q,r) 4 \pi^2 t^2 \approx \lambda^2\frac{0.0016}{m} 4 \pi^2 t^2
\ee
where we have used the following numerical result that will be explored in the next section.
\be\label{sum B}
\sum_{k=1\in\mathbb{Z}_{\text{odd}}}^{2m-1}~B^{11\to 11}_{k, k}(m;m/3,m/3,m/3)\approx \frac{0.0016}{m}
\ee
We expect that there is a smooth connection between the linear $t$ behavior (\ref{amp linear}) and quadratic $t$ behavior (\ref{amp t2}) at $\frac{1}{m^2}t \sim \frac{1}{m}t^2$, which is
\be
t\sim O(m^{-1})
\ee
In summary, the amplitude is
\be\label{amp t and t2}
A_m^{11\rightarrow 11}(t) \approx \begin{cases}
\lambda^2\,\frac{0.0016}{m} 4 \pi^2 t^2 & 0\leq t \lesssim O(m^{-1})\\
\lambda^2\,\frac{\pi}{18 m^2} t & O(m^{-1})\lesssim t \lesssim \pi-O(m^{-1})
\end{cases}
\ee
From (\ref{amp B}), we find that the amplitude has periodicity $2\pi$, which can also be seen from fig.\,\ref{fig_amp}. The linear behavior is only valid for some regions. When the time scale is much larger than $1/m$, the quadratic behavior can be neglected and the amplitude can be approximated as
\be\label{amp saw}
A_m^{11\rightarrow 11}(t) \approx \lambda^2\,\frac{c}{ m^2} [t]_{\text{saw-like}, 2\pi}
\ee
where $c\approx \frac{\pi}{18}$.
In the next section, we will study how this linear behavior in the region $(0,\pi)$ emerges from the sum in (\ref{amp B}).

\subsection{Properties of $B^{ij\to i'j'}_{k, k}$}\label{sec B}

To better understand the properties of the amplitude in the large $m$ limit, in this section we will study the properties of the coefficient $B^{ij\to i'j'}_{k, k}$ in the amplitude (\ref{amp B}). The numerical results of $B^{11\to 11}_{k, k}(m;m/3,m/3,m/3)$ for $m=18, 24, 30, 36$ are given in appendix \ref{B values}.

In fig.\,\ref{fig_B}, we plot $m^2 B^{11\to 11}_{k, k}$ as a function of $m$ for different $m$. We can see that for different $m$ the plots follow the same shape. As $m$ becomes larger, more points fit into this curve. We expect that this is true in the limit $m\rightarrow \infty$.

Thus as $m$ becomes larger, more and more points fit into the small $k$ region, say the region $m\lesssim m/6$. In this region the $m^2 B^{11\to 11}_{k, k}$ can be approximated as a constant. From the right panel of fig.\,\ref{fig_B}, we find that the constant is
\be
m^2 B^{11\to 11}_{k, k}\approx 0.0028 ~~~~~~~ k \ll m ~~~~~~m\rightarrow \infty
\ee
\begin{figure}
\centering
        \includegraphics[width=5cm]{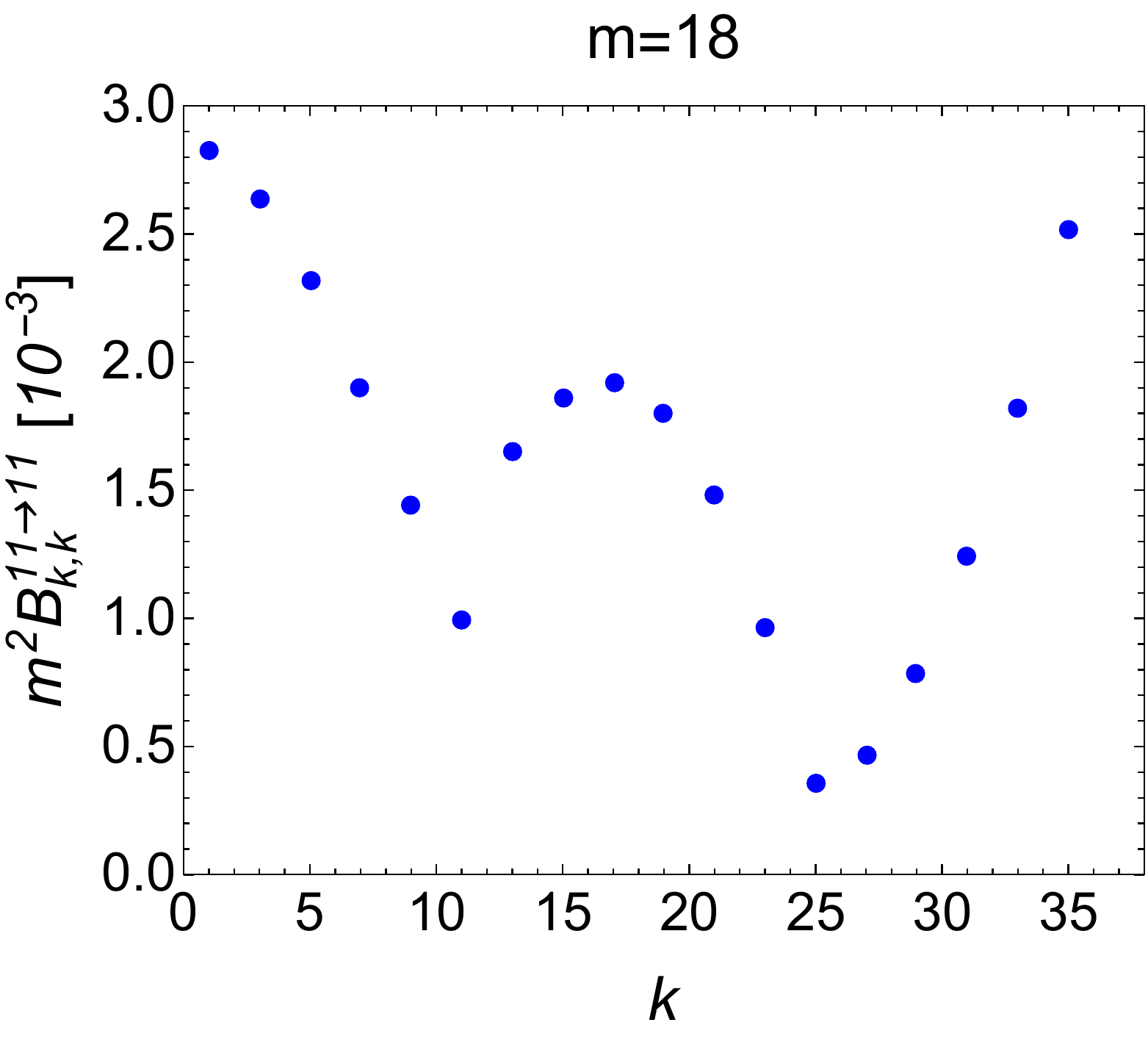}\quad
       \includegraphics[width=5cm]{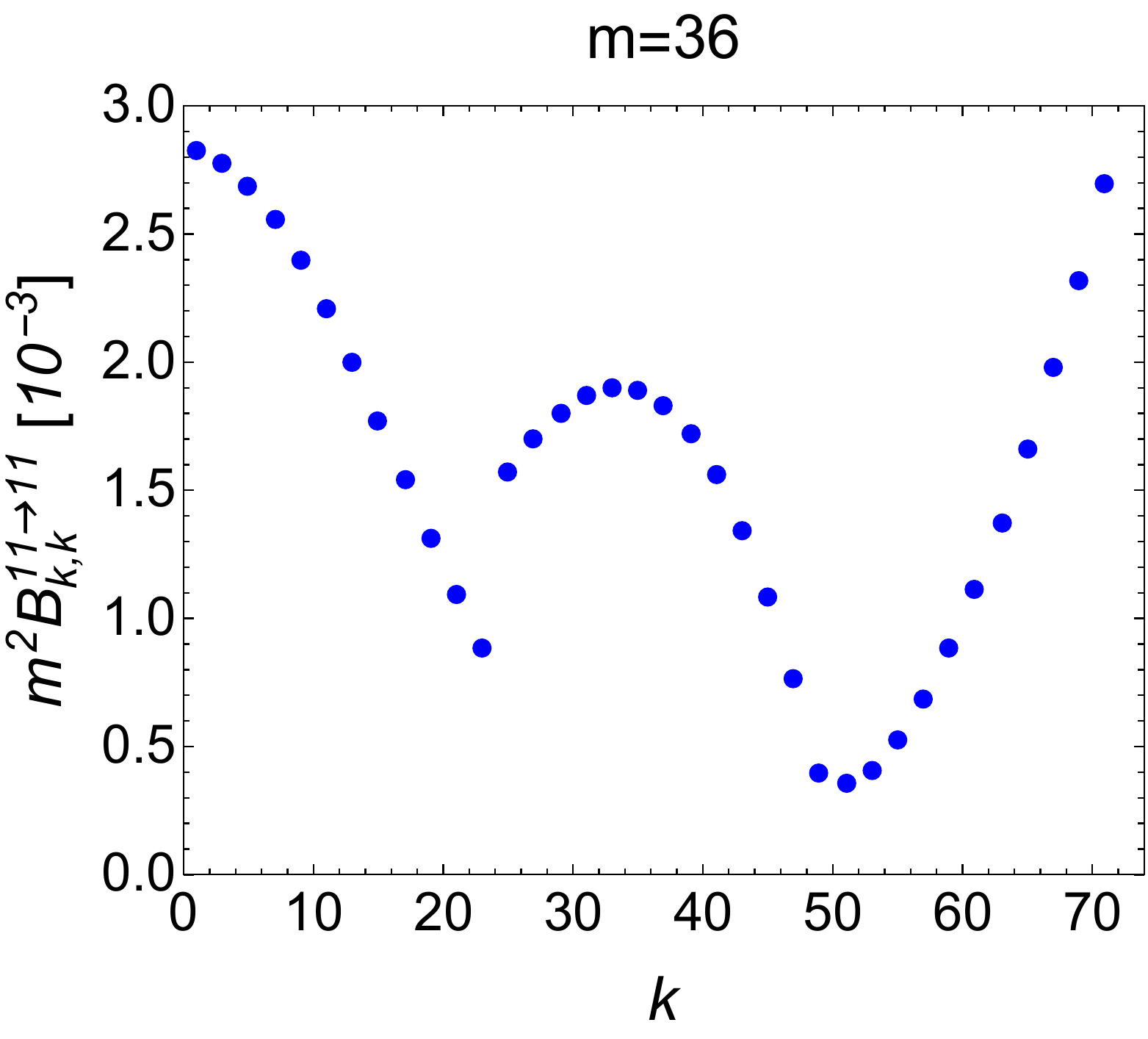}\quad
       \includegraphics[width=5cm]{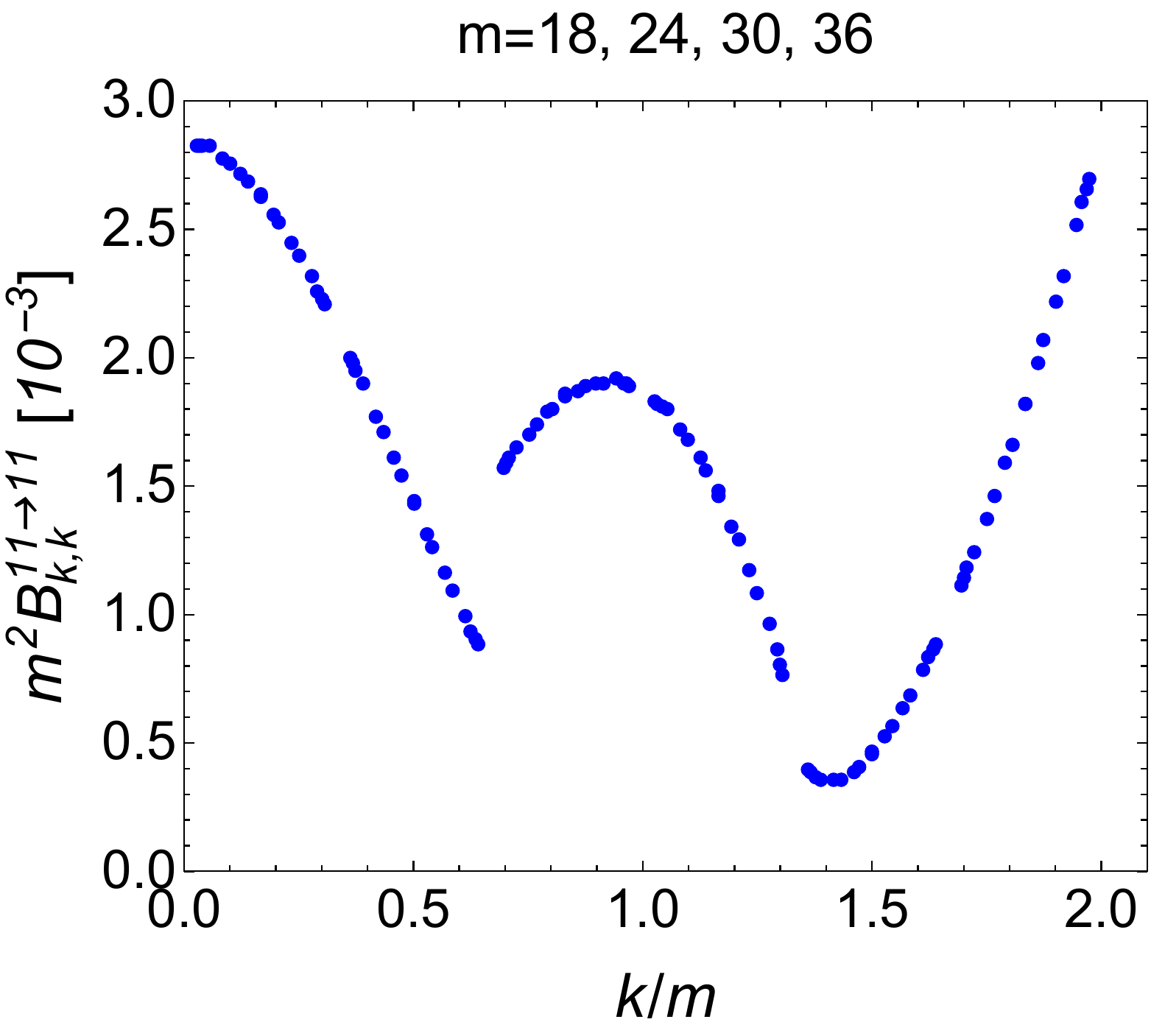}
\caption{$m^2 B^{11\to 11}_{k, k}$ as a function of $m$ for $m=18$ (Left) and $m=36$ (Middle).
$m^2 B^{11\to 11}_{k, k}$ as a function of $\frac{k}{m}$ for $m=18,24,30,36$ (Right).}
\label{fig_B}
\end{figure}
In this approximation, the amplitude $A^{11\rightarrow 11}_m(t)$ (\ref{amp B}) becomes
\bea\label{amp linear approx}
A_m^{11\rightarrow 11}(t)&\approx&\lambda^2 \sum_{k=1\in\mathbb{Z}_{\text{odd}}}^{m/6}\frac{0.0028}{m^2} {16 \pi^2 \over k ^2} 
    \sin^2(  {kt\over 2})\nn
&\approx& \lambda^2 \frac{0.0056 \pi^3}{m^2} [t]_{\text{ saw-like}, 2\pi} ~\approx~ \lambda^2 \frac{\pi}{18 m^2} [t]_{\text{ saw-like}, 2\pi}
\eea
In the large $m$ limit, the upper limit of the summation becomes $m/6\rightarrow \infty$. In the second step we have used
\be\label{linear}
\sum_{k= 1\in\mathbb{Z}_{\text{odd}}}^{\infty}~~~ {1 \over k ^2} 
    \sin^2\left(  {kt\over 2}\right) = \begin{cases}
\frac{\pi}{8}t & 0\leq t \leq \pi\\
\frac{\pi}{8}(2\pi-t) & \pi\leq t \leq 2\pi
\end{cases}
\equiv \frac{\pi}{8}[t]_{\text{ saw-like},2\pi}
\ee
where we sum over all positive odd numbers $k$. 

The (\ref{linear}) explains how the linear region emerges from a sum of periodic functions. Because $\sin^2(  {kt\over 2})$ is bounded, the factor $1/k^2$ makes the sum converge to a linear function in $t$ quickly. From fig.\,\ref{fig_B}, we see that the $B^{11\to 11}_{k, k}$ is bounded as a function of $k$. Thus due to the factor $1/k^2$, which decays at large $k$, the amplitude $A^{11\rightarrow 11}_m(t)$ (\ref{amp B}) can be well approximated by using $k$ with small values, say $k\lesssim m/6$. 

Now let us verify eq.\,(\ref{sum B}). From fig.\,\ref{fig_B}, the curves of $n^2 B^{11\to 11}_{k, k}$ as a function of $k$ for different $m$ are the same. There are $m$ points of $m$ fitting into this curve. Thus the sum of $m^2 B^{11\to 11}_{k, k}$ over $k$ is proportional to $m$, which gives
\be\label{sum B pro}
\sum_{m=1\in\mathbb{Z}_{\text{odd}}}^{2n-1}~~~ B^{11\to 11}_{k, k}(m;m/3,m/3,m/3)\propto \frac{1}{m}
\ee
The proportional constant can be fixed by summing the numerical results in Appendix \ref{B values}, which gives (\ref{sum B}).

\subsection{Amplitudes involving other copies}\label{sec N}

In section \ref{sec amp}, we studied the amplitude $A_m^{11\rightarrow 11}(t)$, where the initial and final bosonic modes are placed on copy 1 in the splitting process (\ref{process}). In this section, we will study the general amplitudes $A_m^{ij\rightarrow i'j'}(t)$, where the $i,j,i',j'$ can be $1,2$.

By studying the numerical results, we find that 
\be\label{amp copy}
A_m^{ij\rightarrow i'j'} = (-1)^{i+j+i'+j'}A_m^{11\rightarrow 11}~~~~~~~i,j,i',j'=1,2
\ee
Every time we change $1$ to $2$ in the amplitude $A_m^{11\rightarrow 11}$ we get a minus sign. 
It is convenient to use the base where the splitting processes are
\bea\label{process 1pm2}
|\Phi_0;1\pm 2,1 \pm 2\rangle&=&(\alpha^{(1)}_{--,-m}\pm \alpha^{(2)}_{--,-m})(\bar \alpha^{(1)}_{++,-m}\pm\bar \alpha^{(2)}_{++,-m}) |\Omega_{1,2}\rangle\nn
\rightarrow |\Phi_f;1\pm 2,1\pm 2\rangle&=&(\alpha^{(1)}_{--,-p} \alpha^{(1)}_{++,-q} \alpha^{(1)}_{--,-r}\pm\alpha^{(2)}_{--,-p} \alpha^{(2)}_{++,-q} \alpha^{(2)}_{--,-r}) \nn
&& (\bar \alpha^{(1)}_{++,-p}  \bar \alpha^{(1)}_{--,-q}  \bar \alpha^{(1)}_{++,-r}\pm \bar \alpha^{(2)}_{++,-p}  \bar \alpha^{(2)}_{--,-q}  \bar \alpha^{(2)}_{++,-r}) |\Omega_{1,2}\rangle
\eea
In this new basis, we will label the amplitude as $A_m^{ij\rightarrow i'j'}(t)$ with $i,j,i',j'=1+2,1-2$.
The mode labeled by $1+2$ is symmetric between the two copies. The mode labeled by $1-2$ is antisymmetric between the two copies.
From (\ref{amp copy}), we find that amplitudes involving symmetric mode are zero. Thus the only nontrivial amplitude is the $A_m^{1-2,1-2\rightarrow 1-2,1-2}$ in (\ref{process 1pm2}).

Recall that $\alpha^{(1)}_{--,-m}+ \alpha^{(2)}_{--,-m}$ is a `global' mode defined in (\ref{global}), which is a symmetry of the system of two component strings. Thus it can not be split by the deformation operator, which is consistent with the result that the amplitude is zero.

From (\ref{amp copy}), we find that for the case of two singly wound strings, there are collisions between the left and right movers on the same copy and on different copies. The resulting left and right movers can be on the same and different copies. 
The amplitudes have the same magnitude.

\section{Properties of the splitting process}\label{sec operator growth}

\subsection{Large $N$ limit}

In section \ref{sec numerical}, we found that for the case of two singly wound strings the amplitude of the splitting is
\be\label{amp linear 2}
A_m^{11\rightarrow 11}(t) \approx \lambda^2 \frac{c}{m^2} [t]_{\text{saw-like},2\pi}
\ee
where $c\approx \frac{\pi}{18}$. Other processes involving copy 2 can be obtained from (\ref{amp copy}). They all have the same magnitude. Thus if we start with excitations on copy 1, after the splitting process, the resulting excitations can easily `move' to copy 2.  In this section, we will study the case of $N$ singly wound strings. We will find that if we start with excitations on a particular copy, the resulting excitations will stay on this copy in the large $N$ limit.

Let us first consider the amplitude $\tilde{A}_m^{11\rightarrow 11}(t)$ where the initial and final excitation are on copy 1 and the remaining $N-1$ copies are in the NS vacuum. We use $\tilde{A}$ to label the amplitude when there are $N$ singly wound strings. The two deformations can twist and untwist copy 1 with any of the remaining $N-1$ strings. The amplitude is enhanced by a factor of $N-1$
\be\label{amp linear N}
\tilde{A}_m^{11\rightarrow 11}(t) \approx (N-1)\lambda^2 \frac{c}{m^2} [t]_{\text{saw-like},2\pi} \approx g^2 \frac{c}{m^2} [t]_{\text{saw-like},2\pi}
\ee
where the t' Hooft coupling is defined as $g=\lambda N^{1/2}$. 

If the resulting left and right movers are not on the same copy, for example consider
\be
\tilde{A}_m^{11\rightarrow 1j'}(t) \approx \lambda^2 \frac{c}{m^2} [t]_{\text{saw-like},2\pi} ~~~~~~~j'= 2,3,\dots,N
\ee
Here we do not have the enhancement factor $N$ because different values of $j'$ correspond to different final states.
In the large $N$ limit, summing over all possible values of $j'$, we have
\be
\sum_{j'=2}^{N}|\tilde{A}_m^{11\rightarrow 1j'}(t)|^2 \approx |\lambda^2 \frac{c}{m^2} [t]_{\text{saw-like},2\pi}|^2 N \ll |\tilde{A}_m^{11\rightarrow 11}(t)|^2
\label{11 to 1l}
\ee
Note that for the $11\to11$ process the probability comes with a factor of $N^2$ whereas for the $11\to 1j'$ the probability comes with a singly power of $N$. Thus the resulting left and right movers prefer to stay on the same copy as the initial state. In the large $N$ limit, the dominate splitting process will not spread the excitations onto more and more copies and is given by
\be\label{amp dominate}
\tilde A_m^{ii\rightarrow ii}(t) \approx g^2 \frac{c}{m^2} [t]_{\text{saw-like},2\pi}
\ee
where $i=1,\dots,N$ labels one of the singly wound component strings.

\subsection{Effective local operator}\label{sec local operator}

We have studied the splitting process in the previous sections. 
In this section, we will interpret the amplitude (\ref{amp dominate}) in the region $(0,\pi)$ 
\be\label{A local}
\tilde{A}^{ii\rightarrow ii}_{m}(t)\propto \frac{1}{m^2}t
\ee
as a result of an effective local operator. We also discuss its consequence on operator growth. 

Notice that in the amplitude (\ref{A integral}) there are two time integrals for the time of the two deformation operators. The linear $t$ behavior in (\ref{A local}) indicates that the contribution when the two deformations are separated far away is small. The two deformation operators are bound together and thus there is only one time integral left. However, not all linear $t$ behavior comes from local operators. In the following, we will show that the $1/m^2$ dependence in the amplitude (\ref{A local})  is a consequence of a local operator.

Suppose $L$ is the length of a singly wound string. There are two energy scales in the process, the energy scale $1/L$ from the size of the cylinder and the energy scale $E\sim m/L$ of the splitting process $\alpha_{-m}\bar\alpha_{-m}\rightarrow \left(\alpha_{-m/3}\right)^3\left(\bar\alpha_{-m/3}\right)^3$.
The bosonic modes $\alpha_{-m}$ and $\bar\alpha_{-m}$ come from the local operators $\p X$ and $\bar\p X$ as shown in (\ref{modes}).
Thus the following operator in the Lagrangian can cause the splitting process
\be\label{local operator}
g(E) (\p X)^4 (\bar\p X)^4
\ee
where $g(E)$ is the coupling which can depend on the energy scales. The $\p X$ and $\bar\p X$ are local operators both with mass dimension $1$. The interaction (\ref{local operator}) must have total mass dimension $2$ as required by being a term in the Lagrangian in two dimensions.
Thus the coupling $g(E)$ has mass dimension $-6$. If it is a local operator, it should only depend on the energy scale $E\sim m/L$ but not $1/L$
\be
g(E)\sim \frac{g}{E^6}
\ee
Therefore, the operator (\ref{local operator}) will not see the size of the cylinder and cannot involve the energy scale $1/L$, which is an indication of non-locality.
In terms of normalized fields $\frac{1}{\sqrt{E}}\p X$ and $\frac{1}{\sqrt{E}}\bar\p X$, the local operator (\ref{local operator}) can be written schematically as
\be
\frac{g}{E^6}(\p X)^4 (\bar\p X)^4 \sim \frac{g}{E^2} (\frac{1}{\sqrt{E}}\p X)^4 (\frac{1}{\sqrt{E}}\bar\p X)^4
\ee
Thus the amplitude for the splitting process from the local operator, in which the modes are all normalized, should be
\be
\frac{g}{m^2}t
\ee
This explains the $1/m^2$ and $t$ behavior in (\ref{A local}). Thus the amplitude can result from an effective local operator like (\ref{local operator}).

The effective local operator breaks down at very early times $0\leq t \lesssim O(m^{-1})$, where we have $\tilde{A}^{ii\rightarrow ii}_{m}(t)\sim t^2$ from (\ref{amp t and t2}). The connection between the $t$ and the $t^2$ behavior is at $t\sim O(m^{-1})$. Thus the two deformations form a bound state of size $O(m^{-1})$. When the time is shorter than $O(m^{-1})$, they can move freely and result in $t^2$ behavior. When the time is longer than $O(m^{-1})$, they form a bound state and result in $t$ behavior. This can also be understood from the properties of $B^{ij\to i'j'}_{k, k}$ in the amplitude (\ref{amp B}), where  $-k/2$ is the dimension of the intermediate states relative to the initial energy. In section \ref{sec B}, $B^{ij\to i'j'}_{k, k}$ has the same magnitude for $|m|\lesssim O(m)$. The uncertainty in energy is $O(m)$, which indicates that the uncertainty in time is $O(m^{-1})$. Thus the two deformation operators are bounded with a size $O(m^{-1})$ in the time direction.

\subsection{A freely falling graviton in AdS}

In this section, we will relate the properties of the splitting process to the behavior of a freely falling graviton in the bulk. We will discuss the behavior at early times and late times separately.

\bigskip

(i) Early time behavior, $0\lesssim t\lesssim \pi$: 
Recall that at early times the splitting is caused by an effective local operator.
Consider a wave-packet containing a left mover and a right mover $\alpha_{-m}\bar\alpha_{-m}$ which has a size of about $1/m$. The probability of the splitting should only involve the energy scale $m$ of the wave-packet. Further, the probability is proportional to $g^4$ and $t^2$ since it is proportional to the modulus squared of the amplitude. Because probability is dimensionless, it can be written as
\be
P\sim g^4 m^2 t^2
\ee
Setting $P\sim 1$, the time for splitting can be estimated as
\be
t\sim \frac{1}{g^2 m}
\ee
If we extrapolate it to the strongly coupled region where $g\sim O(1)$, we find the time is about $t\sim 1/m$. During this time, the wavelengths of the modes increase from $1/m$ to $3/m$. Thus the growth is linear in time, which is expected if the splitting is caused by a local operator.
As mentioned in the introduction, linear growth of the wavelength in the CFT corresponds to a free falling graviton in AdS. As the graviton moves deeper in AdS, it becomes more and more redshifted and the wavelengths of the modes within the CFT become longer.

\bigskip

(ii) Late time behavior, $\pi \lesssim t $:
For a free falling graviton in AdS, it will travel along  a geodesic, and by symmetry all points along its path are equivalent. The freely moving graviton in AdS will not split into lower energy gravitons or excite into stringy states.
From eq.\,(\ref{amp B}),
we find that in the CFT the splitting amplitude has periodicity $2\pi$. The effect of the perturbation generates oscillations in the splitting amplitude, but does not lead to a secular term where the amplitude continues to grow \cite{hm}. Therefore, the CFT computation to this order agrees with the expectation from gravity.
Furthermore, the amplitude is zero at $t=2\pi$. In the weakly coupled CFT, the initial bosons will not split to lower energy bosons if the time is $2\pi$. From the bulk point of view, the infalling graviton takes time $2\pi$ to travel across AdS and back to the initial state. The decreasing region in the amplitude shown in fig.\,\ref{fig_amp} may be an indication of this return. 

\section{Discussion} \label{sec discussion}

Here we studied a freely falling graviton propagating in AdS in the context of the D1D5 CFT. To do this we computed transition amplitudes for one left and right moving boson to split into three left moving and right moving bosons at second order in the deformation operator as depicted in fig.\,\ref{fig_splitting}. 

For an initial left and right mover each with dimension $m$, which correspond to a total energy of $2m$, we showed that the amplitude oscillates with a period of $2\pi$. Within each period the amplitude rises to a maximum at a multiple of $\pi$. Before time $t\sim O({1\over m})$ the amplitude grows like $t^2$ meaning that each deformation operator moves freely. After a time $t\sim O({1\over m})$ the amplitude switches to a linear growth in $t$. This signifies that the two deformation operators bind together and  act as an effective local operator of size $\sim O({1\over m})$. This linear behavior becomes more and more pronounced as we increase the value of $m$. This linear growth corresponds to an infalling graviton becoming redshifted in AdS. For a discussion about redshift and thermalization in the context of fuzzballs, see section 8 of \cite{Carson:2016cjj}.

One may wonder why the amplitude is periodic and doesn't continue to grow in $t$. 
We have shown that the following two properties of the intermediate states between the two deformation operators are essential. First, there is no intermediate states with exactly the same left and right dimensions as the initial state. 
Second, the intermediate states with energies increasing and decreasing with same amount compared to the initial energy should have the same amplitude as shown in (\ref{B sym}). The above properties of intermediate states may be found by studying the effect of one deformation operator \cite{Avery:2010er}. We hope to return to this in a future work. 
The periodic behavior of the amplitude in the CFT is consistent with the gravity dual. In AdS, a single graviton which is sent in from the boundary should propagate freely through the bulk to the other side and  back again to its starting point. We don't expect that it will split into lower energy gravitons or become excited into stringy states, which is a signal of thermalization \cite{Carson:2016cjj}.

In previous work \cite{hm,dissertation} we found the oscillatory behavior for the one to three splitting process for some low energies. In this paper, we show that there is oscillatory behavior for large energies and analyze it in detail. In \cite{hm,dissertation} we also computed a two to four process which was shown to grow like $t^2$, which we argued was a preliminary signal of thermalization. This would correspond to two gravitons colliding in the gravity dual. In \cite{mw}, the tidal force on an infalling graviton  is studied in the $(1,0,n)$ superstratum geometry. The infalling graviton would become tidally excited into string states. In the CFT dual, the state corresponding to the $(1,0,n)$ superstratum is in the Ramond sector. It has extra left moving modes with nonzero energy and right moving modes with zero energy compared to the case we studied in this paper. If we consider the splitting of one left and right mover in this CFT background, we expect that the left and right mover will collide with extra left and right movers in the background. This could result in a growing term like $t^2$ as shown in (\ref{t^2}) and may explain the tidal force in \cite{mw}. We hope to return to this in a future work.

\section{Acknowledgements}
We would like to thank Samir Mathur for helpful discussions.
The work of B.G. is supported by DOE Grant DE-SC0011726. The work of S.H. is supported by the ERC Grant 787320 - QBH Structure.

\appendix

\section{$B^{11\to 11}_{k, k}(m;m/3,m/3,m/3)$}\label{B values}
In this appendix we tabulate the coefficients, $B^{11\to 11}_{k, k}(m;m/3,m/3,m/3)$ for initial energies $m=18,24,30,36$. Figs. \ref{fig_amp}, \ref{fig_B} were made using exact values of rational numbers but for brevity we present them here in numerical form. We keep an appropriate level of precision such that there is no distinguishable difference between the plots using the exact values and plots using the approximate values here. 

\begin{table}[!ht]
\centering
\begin{tabular}{||c|c|c|c|c||}
\hline
k & $B^{11\to 11}_{k, k}(18;6,6,6)$&$B^{11\to 11}_{k, k}(24;8,8,8)$&$B^{11\to 11}_{k, k}(30;10,10,10) $ & $B^{11\to 11}_{k, k}(36;12,12,12)$ \\
\hline
 1 & $8.711\times10^{-6}$&  $4.903\times10^{-6}$ & $3.138\times10^{-6}$ &$2.180\times10^{-6}$\\
 \hline
 3 &$ 8.147\times10^{-6}$ &  $4.722\times10^{-6}$ & $3.064\times10^{-6}$&$ 2.144\times10^{-6}$ \\
 \hline
 5 & $7.142\times10^{-6}$ &  $4.382\times10^{-6}$& $2.920\times 10^{-6}$ & $2.073\times10^{-6}$  \\
 \hline
 7 & $5.857\times10^{-6}$ &  $3.922\times10^{-6}$ & $2.719\times 10^{-6}$ & $1.973\times10^{-6}$  \\
 \hline
 9 & $4.446\times10^{-6}$&  $3.379\times10^{-6}$ & $2.474\times 10^{-6}$ &$1.847\times10^{-6}$  \\
 \hline
 11 & $3.060\times10^{-6}$& $2.791\times10^{-6}$ & $2.196\times 10^{-6}$& $1.701 \times 10^{-6}$\\
 \hline
 13 & $5.103\times10^{-6}$& $2.194\times10^{-6}$ & $1.898\times 10^{-6}$ & $1.540\times10^{-6}$\\
 \hline
 15 & $5.737\times10^{-6}$& $1.622\times10^{-6}$ & $1.592\times 10^{-6}$ & $1.369\times10^{-6}$ \\
 \hline
 17 & $5.922\times10^{-6}$& $2.799\times10^{-6}$ & $1.289\times 10^{-6}$ & $1.193\times10^{-6}$  \\
 \hline
 19 & $5.552\times10^{-6}$& $3.103\times10^{-6}$ & $1.001\times 10^{-6}$ &  $1.016\times10^{-6}$ \\
 \hline 
 21 & $4.567\times10^{-6}$& $3.283\times10^{-6}$ & $1.764\times 10^{-6}$ &  $8.430\times10^{-7}$ \\
 \hline
 23 & $2.968\times10^{-6}$&$ 3.305\times10^{-6}$ & $1.930\times 10^{-6}$ & $6.787\times10^{-7}$  \\
 \hline
 25 & $1.092\times10^{-6}$& $3.148\times10^{-6}$ & $2.050\times 10^{-6}$& $1.213\times10^{-6}$\\
 \hline
 27 & $1.430\times10^{-6}$& $2.798\times10^{-6}$ & $2.112\times 10^{-6}$ & $1.313\times10^{-6}$\\
 \hline
 29 & $2.428\times10^{-6}$& $2.246\times10^{-6}$ & $2.106\times 10^{-6}$&  $1.392\times10^{-6}$ \\
 \hline
 31 & $3.851\times10^{-6}$& $1.502\times10^{-6}$ & $2.025\times 10^{-6}$ & $1.445\times10^{-6}$\\
 \hline
 33 & $5.632\times10^{-6}$& $6.427\times10^{-7}$ & $1.865\times 10^{-6}$ & $1.468\times10^{-6}$ \\
 \hline
 35 & $7.757\times10^{-6}$& $6.728\times10^{-7}$ & $1.624\times 10^{-6}$ & $1.457\times10^{-6}$\\
 \hline
 37& --- &  $9.771\times10^{-7}$ & $1.299\times 10^{-6}$ & $1.411\times10^{-6}$ \\
 \hline
 39&--- &  $1.448\times10^{-6}$ & $8.960\times 10^{-7}$ & $1.326\times10^{-6}$\\
 \hline
 41& ---&  $2.048\times10^{-6}$ & $4.292\times 10^{-7}$&  $1.202\times10^{-6}$\\
 \hline
 43&--- &  $2.761\times10^{-6}$ & $4.004\times 10^{-7}$ & $1.038\times10^{-6}$\\
 \hline
 45& ---&  $3.584\times10^{-6}$ & $5.108\times 10^{-7}$ & $8.340\times10^{-7}$\\
 \hline
 47&--- &  $4.517\times10^{-6}$ & $7.044\times 10^{-7}$ & $5.916\times10^{-7}$ \\
 \hline
 49& ---& --- & $9.597\times 10^{-7}$&  $3.089\times10^{-7}$ \\
 \hline
51 &--- & --- & $1.267\times 10^{-6}$&  $2.715\times10^{-7}$ \\
 \hline
53 &--- & --- & $1.621\times 10^{-6}$& $3.141\times10^{-7}$  \\
 \hline
55 &--- &--- & $2.020\times 10^{-6}$& $4.043\times10^{-7}$ \\
 \hline
57 &--- &--- & $2.463\times 10^{-6}$& $5.292\times10^{-7}$ \\
 \hline
59 &--- &--- & $2.952\times 10^{-6}$&   $6.822\times10^{-7}$ \\
\hline
61 &--- &--- &--- &  $8.598\times10^{-7}$\\
\hline
63 &--- &--- &--- &  $1.060\times10^{-6}$\\
\hline
65 & ---& ---& ---&  $1.282\times10^{-6}$\\
\hline
67 & ---& ---&--- &  $1.526\times10^{-6}$\\
\hline
69 &--- &--- &--- &  $1.791\times10^{-6}$\\
\hline
71 & ---&--- & ---&  $2.078\times10^{-6}$\\
 \hline
\end{tabular}
\caption{Tabulation of coefficients $B^{11\to 11}_{k, k}(m;m/3,m/3,m/3)$ for various values of initial energy $m$.}
\label{table of Bs}
\end{table}
\clearpage

\end{document}